\documentclass[conference]{IEEEtran}
\IEEEoverridecommandlockouts
\usepackage[left=15.7mm,right=15.7mm,top=19mm,bottom=24.5mm]{geometry}
\usepackage[T1]{fontenc}
\usepackage[utf8]{inputenc}
\usepackage{cite}
\usepackage{amsmath,amssymb,amsfonts}
\usepackage{algorithmic}
\usepackage{algorithm}
\usepackage{graphicx}
\usepackage{textcomp}
\usepackage{tabularx}
\usepackage{xcolor}
\usepackage[normalem]{ulem}
\usepackage{bm}

\def\BibTeX{{\rm B\kern-.05em{\sc i\kern-.025em b}\kern-.08em
    T\kern-.1667em\lower.7ex\hbox{E}\kern-.125emX}}
\begin{document}

\title{Mitigating Evasion Attacks in Fog Computing Resource Provisioning Through Proactive Hardening
\thanks{The presented work has been funded in whole by the National Science Centre, Poland, within the project no. 2023/05/Y/ST7/00002 PASSIONATE within the CHIST-ERA programme. For the purpose of Open Access, the author has applied a CC-BY public copyright licence to any Author Accepted Manuscript (AAM) version arising from this submission.}}

\author{\IEEEauthorblockN{Younes Salmi}
\IEEEauthorblockA{\textit{Institute of Radiocommunications} \\
\textit{Poznan University of Technology}\\
Poznań, Poland \\
younes.salmi@put.poznan.pl}
\and
\IEEEauthorblockN{Hanna Bogucka}
\IEEEauthorblockA{\textit{Institute of Radiocommunications} \\
\textit{Poznan University of Technology}\\
Poznań, Poland \\
hanna.bogucka@put.poznan.pl}
}

\maketitle

\begin{abstract}
This paper investigates the susceptibility to model integrity attacks that overload virtual machines assigned by the \emph{k-means} algorithm used for resource provisioning in fog networks. 
The considered k-means algorithm runs two phases iteratively: offline clustering to form clusters of requested workload  
and online classification of new incoming requests into offline-created clusters.
First, we consider an evasion attack against the classifier in the online phase.
A threat actor launches an exploratory attack using query-based reverse engineering to discover the Machine Learning (ML) model (the clustering scheme).
Then, a passive causative (evasion) attack is triggered in the offline phase.
To defend the model, we suggest a proactive method using adversarial training to introduce attack robustness into the classifier. 
Our results show that our mitigation technique effectively maintains the stability of the resource provisioning system against attacks.
\end{abstract}


\section{Introduction}
\label{sec:into}
The fog computing market is poised to grow exponentially, reaching \$6.38 Billion by 2032 \cite{skyquesttComputingMarket}.
This growth reflects
the importance of this computational paradigm that brings computational resources closer to the data source. Fog computing networks allow for the reduction of latencies and traffic congestion compared to cloud computing \cite{fog2C}.
Fog nodes are usually more computationally and energy-constrained than cloud servers.
Moreover, a dynamically changing fog environment requires dynamic resource management and is one of the challenging problems to take into account in fog network management \cite{jamil2022resource}.

In the literature, extensive research has been documented covering the applications of resource management, such as resource scheduling, task offloading, load balancing, and resource provisioning \cite{Kopras2022TCOM, ghobaei2020resource}. 
Virtualization techniques such as Virtual Machine (VM) placement are used for Resource Provisioning Systems (RPSs) to match the user demands \cite{santos2019resource}. 
However, with the growing number of users of the computing platform, static RPSs can fail with heterogeneous dynamic workloads, resulting in resource wastage or insufficient provisioning \cite{shakarami2022resource}.
This calls for the use of sophisticated techniques for fog resource management, such as Machine Learning (ML) \cite{walia2023ai}.

Supervised learning techniques such as \emph{Autoregressive Integrated Moving Average} enable proactive resource management, reducing \emph{Service Level Agreement} violations \cite{ARIMA}.
Semi-supervised Reinforcement Learning (RL) approaches ensure dynamic autoscaling and load balancing in real-time systems with long-term rewards \cite{asghari2021task}.
Supervised techniques are based on high-quality labeled data and retraining, which may not be feasible in a dynamic system \cite{van2020survey}.
Moreover, RL methods may be computationally complex, have slow convergence, and require the effective design of reward functions \cite{hu2024review}. 

For realistic \textbf{C}ommunication and \textbf{C}omputing (2C) resource management in a fog, unsupervised learning, such as clustering, offers a more scalable approach by discovering hidden patterns in unlabeled data \cite{elastic}. 
In contrast to supervised methods, these techniques excel in dynamic computing platforms when workloads are heterogeneous and unpredictable.
Moreover, they quickly adapt to changing patterns and distributions of data with less computational overhead compared to RL methods.
As a result, techniques such as clustering \cite{alnoman2020machine, ullah2020task, mtshali2019k, cheng2019balanced} enable resource allocation based on workload grouping.
Regardless, these approaches are poorly tested for large-scale mobile communication networks and their multi-objective management.    

ML-based RPS, like all ML methods, are susceptible to vulnerabilities, including model extraction, data manipulation, and adversarial examples \cite{nazari2023adversarial}. 
These threats expand the physical and digital attack surfaces, leading to increased cybersecurity risks.
For example, threat actors may exploit vulnerabilities in RPS by launching adversarial examples to manipulate the boundaries of ML classifiers.
In \cite{nazari2023adversarial, makrani2021security}, the authors aimed to bypass security measures by manipulating RPS algorithms to co-locate malicious VM with a target victim's VM. 
In 
\cite{nazari2023adversarial}, the model is attacked by exploratory attack, namely reverse engineering, to discover the decision-making process of the ML-based RPS.
Then, adversarial examples are crafted using a Fake Trace Generator (FTG) to manipulate the input examples.
Beyond direct data manipulation, physical attacks such as RF jamming \cite{salmi2023security} and data manipulation using Python scripts \cite{younes2023attacks, salmi2024poisoning} further demonstrate the diverse methods threat actors use.

The results of these works show that ML-based RPS are vulnerable to adversarial examples, and data errors may occur due to malicious manipulation of decision regions.
Thus, securing ML-based RPS models is a critical challenge to ensure fair and reliable resource management.
Several hardening (or defense) techniques have been proposed. 
In \cite{salmi2024poisoning}, for example, an Anomaly Detection (AD) method was proposed to filter out poisoned samples injected into the data set. 
In \cite{Wasilewska2023ComMag}, poisoned models are discovered (as anomalies) based on their statistical comparison.
However, AD is considered a non-real-time and reactive defense strategy, employing computationally complex statistical tools reactive to attacks.
Possible proactive hardenings have been recommended in \cite{nazari2023adversarial}, namely the model and training modification approaches.
These techniques should ensure the classifier's resilience to adversarial examples using tools such as adversarial training.
However, none of them has been implemented to combat attacks against ML-based RPS. 

Given the challenges discussed above, this paper investigates the attack surface of ML-based RPS with on-demand VM allocation using workload clustering and classification \cite{elastic, alnoman2020machine, ullah2020task, mtshali2019k, cheng2019balanced}, as well as proactive hardening of this RPS against attacks. The considered RPS aims at assigning the workload (users' demands) to fog network computational resources (VMs) using the \emph{k-means} algorithm. 
The considered k-means algorithm runs in two phases iteratively: offline clustering to form clusters of workloads (corresponding to VMs and their capabilities) and online classification to classify new incoming workloads into offline-created clusters.
We assume that initially, the ML classifier can be created based on real-time 
demands with minimal complexity, and fast convergence.
To do so, the clustering process is deliberately 
initialized based on a level-based approach.
Furthermore, the classifier is assisted with a re-clustering threshold 
to limit the number of k-means runs. 

Next, we explore a novel multiphase adversarial machine learning (AML) attack launched in the following phases:
\begin{enumerate}
    \item Exploratory attack 
    on the existing (initialized and converged) classifier using reverse engineering to discover the classifier boundaries.
    \item Evasion attack using adversarial examples in the 
    online phase of the workload classification aiming to expand clusters corresponding to resource-constrained VMs.
    \item Causative attack in the 
    offline phase to overload the VM and drop critical workloads.
\end{enumerate}
Thus, this paper extends the discussion in \cite{nazari2023adversarial} by discovering new vulnerabilities of ML-based RPS.
These vulnerabilities are analyzed to determine the security of critical workloads.
Moreover, this paper proposes the proactive hardening mechanism (as an attack mitigation method) using adversarial training \cite{nazari2023adversarial}. 
The proactive hardening aims not only to enhance classifier robustness against evasion attacks but also to ensure resilience to sub-population biases and rare feature deviations. 
By mitigating adversarial examples, this work ensures reliable resource provisioning for critical workloads.

The rest of the paper is organized as follows. 
Section \ref{sec:sys-model} presents the system model and formulates the problem. 
Section \ref{sec:attack} is devoted to the procedures for implementing the MITRE ATLAS attack framework. 
Section \ref{sec:defense} introduces the proactive hardening technique.
Section \ref{sec:sim} depicts the simulation results and their analyses.
Finally, Section \ref{sec:conc} concludes the work.

\begin{figure}[thb]
    \centering
    \includegraphics[width=\columnwidth]{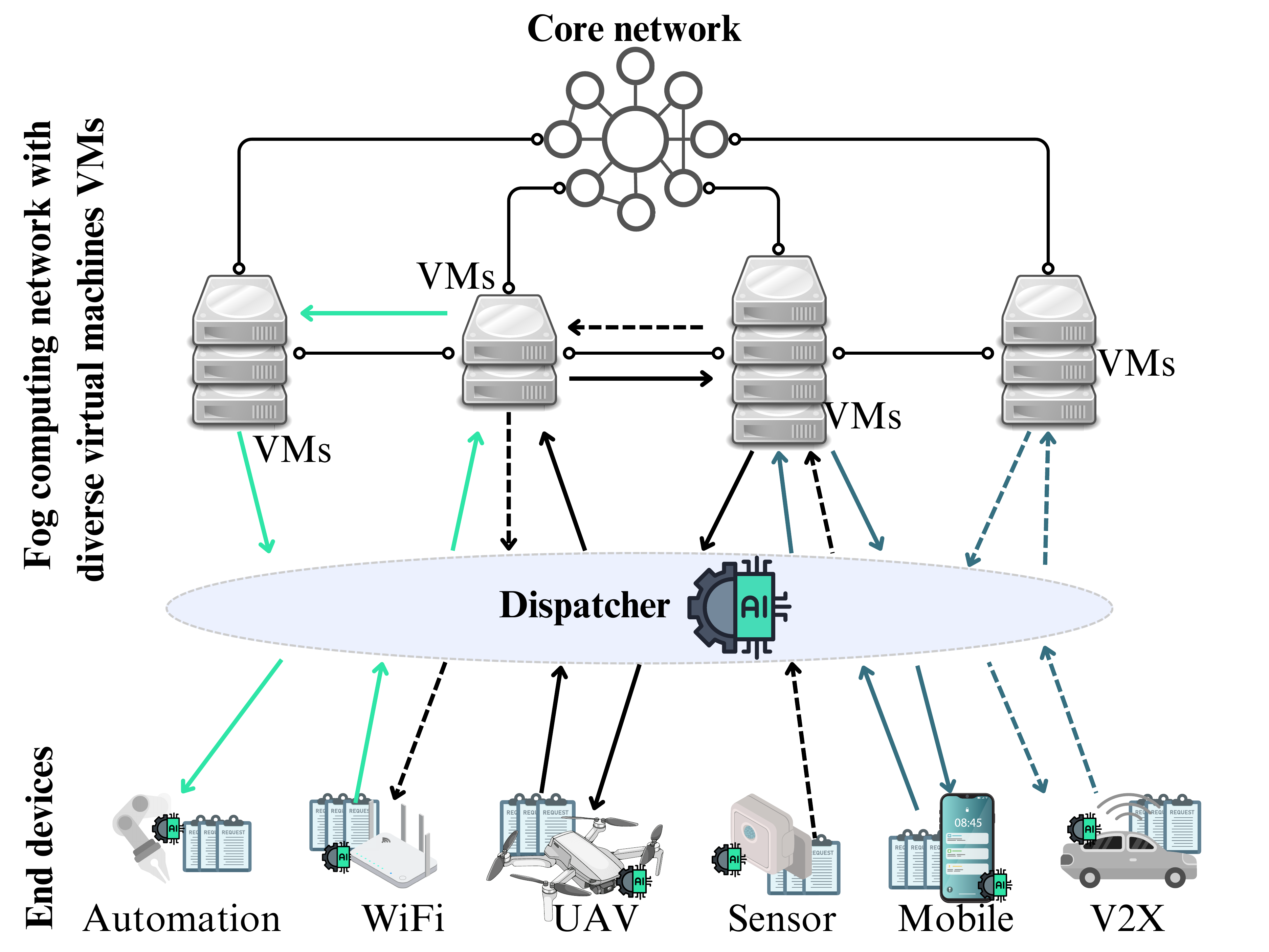}
    \caption{Fog network scenario}
    \label{fig:system-architecture}
\end{figure}
\section{ML-based Resource Provisioning System Scenario}
\label{sec:sys-model}
The considered fog network presented in Fig. \ref{fig:system-architecture} consists of diverse end devices that generate workloads with different 2C demands and the \emph{fog} layer with multiple computing VMs. 
Between these layers, there is an RPS algorithm called \emph{dispatcher}, which directs the requested workload to VMs with the aim of best matching the provisioned resources to 2C requests.

The computational demands of the workload are represented by the required number of Central Processing Unit's (CPU's) Floating-Point Operations (FLOP) $r^{(\text{CPU})}$ and the number of Input-Output (IO) Operations (IOP) $r^{(\text{IO})}$.
Moreover, the End-to-End (E2E) tolerated delay $r^{(\text{E2E})}$ reflects the transmission and service time.  
Finally, the data size requested
to be transmitted $r^{(\text{SIZE})}$ (in bytes) and the tolerated Packet Error Rate (PER) $r^{(\text{PER})}$ represent the communication demands.
Thus, a workload requested by the $m$-th device ($m \in \{1,\ldots,M\}$) is defined by the vector 
$\mathbf{r}_m = [r^{(\text{CPU})}_m,r^{(\text{IO})}_m, r^{(\text{E2E})}_m, r^{(\text{SIZE})}_m, r^{(\text{PER})}_m] \in \mathbb{R}^5$.
The received batch of $M$ 
workloads forms the dataset $\mathbf{R} =\{\mathbf{r}_m\}$.

The generated workloads are served by a number $I$ of VMs, each offering the computational capability $o^{(\text{CPU})}_i$ in FLOP/sec, the read/write capacity $o^{(\text{IO})}_i$ in IOP/sec, a device-to-VM link capacity $o^{(\text{CAP})}_i$ in Bytes/sec, 
and a link PER $o^{(\text{PER})}_i$ (where $1\leq i \leq I$). 
Hence, the serving $i$-th VM is defined by $\mathbf{o}_i = [o^{(\text{CPU})}_i, o^{(\text{IO})}_i, o^{(\text{CAP})}_i, o^{(\text{PER})}_i] \in \mathbb{R}^4$.
The manager's classifier $\mathcal{C}$ clusters the workloads based on their features into the set of $I$ clusters $\mathbf{C} = \{C_i\}$, 
($1\leq i \leq I$) each of them represented by its centroid, 
$C_i \leftrightarrow$  ${\bm{\mu}}_i= [ \mu_i^{(\text{CPU})},\mu_i^{(\text{IO})}, \mu_i^{(\text{E2E})}, \mu_i^{(\text{SIZE})}, \mu_i^{(\text{PER})}] $,   where $\mu_i^{(\text{CPU})}$, $\mu_i^{(\text{IO})}$, $\mu_i^{(\text{E2E})}$,  $\mu_i^{(\text{SIZE})}$, and  $\mu_i^{(\text{PER})}$ are the values of requested FLOP,  IOP E2E delay, data size and PER respectively of $i$-th cluster's centroid, i.e., 
\begin{equation}
     \mathcal{C}\mathrm{:} \;\mathbf{R} \rightarrow \mathbf{C}, \; \mathrm{i.e.,}\; \forall m \, \exists! \,i\mathrm{:} \;\; \mathbf{r}_m  \rightarrow C_i \leftrightarrow \bm{\mu}_i.  
\end{equation}

Then, the manager (dispatcher) identifies VM offers to match the cluster demands, i.e., $ \forall i\in\{1, \ldots, I\}$:

\begin{equation}
        \mathcal{M}\mathrm{:} \;{\bm \mu}_i= \left[ \mu_i^{(\text{CPU})},\mu_i^{(\text{IO})}, \mu_i^{(\text{E2E})}, \mu_i^{(\text{SIZE})}, \mu_i^{(\text{PER})}\right] \leftrightarrow {\mathbf o}_i \;\mathrm{.}
\end{equation}
For ${\mathbf o}_i$ to serve the cluster $C_i$ representative demands defined by its centroid ${\bm \mu}_i$, the average task completion time $t_i$ in cluster $C_i$ must remain shorter than the requested E2E delay $\mu_i^{(\text{E2E})}$. Moreover, PER introduced by the device-to-VM link $e_i$ in that cluster must be lower than $\mu_i^{(\text{PER})}$. 
Ergo, the ${\mathbf o}_i$ offers are chosen to serve clustered demands in a way to satisfy  the following:
\begin{align}
\label{equ:resources-time}
   t_{i} = \frac{\mu_i^{(\text{CPU})}}{\alpha^{(1)}_i \cdot o_i^{(\text{CPU})}} + \frac{\mu_i^{(\text{IO})}}{\alpha^{(2)}_i \cdot o_i^{(\text{IO})}} + \frac{\mu_i^{(\text{SIZE})}}{\alpha^{(3)}_i \cdot o_i^{(\text {CAP})}}& \leq \mu_i^{(\text{E2E})} \\ \label{equ:resources-error}
e_i = o_i^{(\text{PER})} & \leq \mu_i^{(\text{PER})},
\end{align}
where $\alpha^{(1)}_i,\alpha^{(2)}_i, \text{ and } \alpha^{(3)}_i$ are the positive 
factors weighting offered CPU, IOP and link-capacity offers to requested task completion time.
Setting $\alpha_i=\alpha^{(1)}_i=\alpha^{(2)}_i=\alpha^{(3)}_i$ we obtain:
\begin{equation}
\label{equ:alpha}
    \tilde{\alpha}_i = \frac{1}{\mu_i^{(\text{E2E})}}  \left(\frac{\mu_i^{(\text{CPU})}}{o_i^{(\text{CPU})}} + \frac{\mu_i^{(\text{IO})}}{o_i^{(\text{IO})}} + \frac{\mu_i^{(\text{SIZE})}}{o_i^{(\text{CAP})}}\right) \leq \alpha_i \mathrm{,}
\end{equation}
where $\tilde{\alpha}_i$ is the ratio of the offered task completion time over the demanded one (E2E delay demanded) upper bounded by $\alpha_i$. Let us define the ratio of these two by $\Lambda=\frac{\alpha_i}{\tilde{\alpha}_i}$. This metric should be higher for the considered clustering and matching algorithm compared to fixed VM provisioning \cite{alnoman2020machine}. 
To satisfy all the workload demands within $C_i$, the centroid must be moved to the point with the 
lowest $t_i$ and $e_i$. 

For real-time scenarios with new arriving workloads, the re-clustering (offline clustering) is inefficient due to its computational complexity 
$\mathcal{O}(N I M)$, where $N$ represents the number of k-means iterations required for its convergence. 
Subsequently, an online routine is used by assigning a new arriving batch of workloads $\mathbf{R'}$ to the set of clusters $\mathbf{C}$ reducing the complexity to $\mathcal{O}(I M')$, where $M'$ is the cardinality of set $R'$ ($M' = |\mathbf{R'}|$).
The online routine is launched if and only if $\mathbf{R'}$ does not form a new clustering scheme, and this is checked using the Silhouette Score test $\mathcal{S}$ \cite{shahapure2020cluster}.
If $\mathcal{S}$ is above a threshold $\zeta_1$, the online clustering assigns each new workload $\mathbf{r}' \in \mathbf{R'}$ to the nearest centroids.
After that, the online clustering updates centroids if and only if one of $\mathbf{R'}$ workloads is far enough from 
its assigned cluster ($\exists \mathbf{r}'_m \in \mathbf{R'} \quad \text{s.t.} \quad \mathcal{D}(\mathbf{r}'_m, \bm{\mu}_i) \geq \zeta_2$).
If $\mathcal{S}$ is low, the algorithm re-clusters $\mathbf{R}\cup \mathbf{R'}$ using batch clustering that incorporates a combination of previously served workloads in $\mathbf{R}$ and new arrivals in $\mathbf{R'}$. 
This step is essential to handle gradual shifts in workload characteristics.

Furthermore, the offline clustering is intentionally initialized to minimize $N$ and optimize clustering performance.
The algorithm is given two options for a more flexible node capacity virtualization.
The first represents the division of each demand into varying $l$ intervals based on percentile ranks representing the demand levels (e.g. if $l=3$, a demand has three levels, low, moderate, and high).  
Then, the nearest workloads to each range midpoint are set as initial centroids.
This is used to virtualize the nodes in $I$ VMs with varying offer levels, since $\mathbf{R} \in \mathbb{R}^5$, $I = 10^{5\log l}$.
The second option is the inverse of the first, where initially $I$ is defined, and then $l$ is obtained. 
\section{The Multiphase AML Attack}
\label{sec:attack}
This section examines the attack targeting clustering $\mathcal{C}$ to disrupt demanded workload-to-resource matching $\mathcal{M}$. 
The threat actor manipulates a subset of benign online workloads, denoted as $\mathbf{R}_{\mathrm{s}}' \subset \mathbf{R'}$ (evasion attack), and passively poisons the offline data
$\mathbf{R}\cup \mathbf{R'}$ (causative attack). 
This poisoning aims to move benign workloads from higher to lower demand clusters offline.
Consequently, workloads shift from their original to targeted clusters, gradually compromising the provisioning scheme by eroding model integrity.
To improve understanding, we map this attack scenario below to the MITRE ATLAS framework.
 
\begin{figure}[htbp]
    \centering
    \includegraphics[width=\columnwidth]{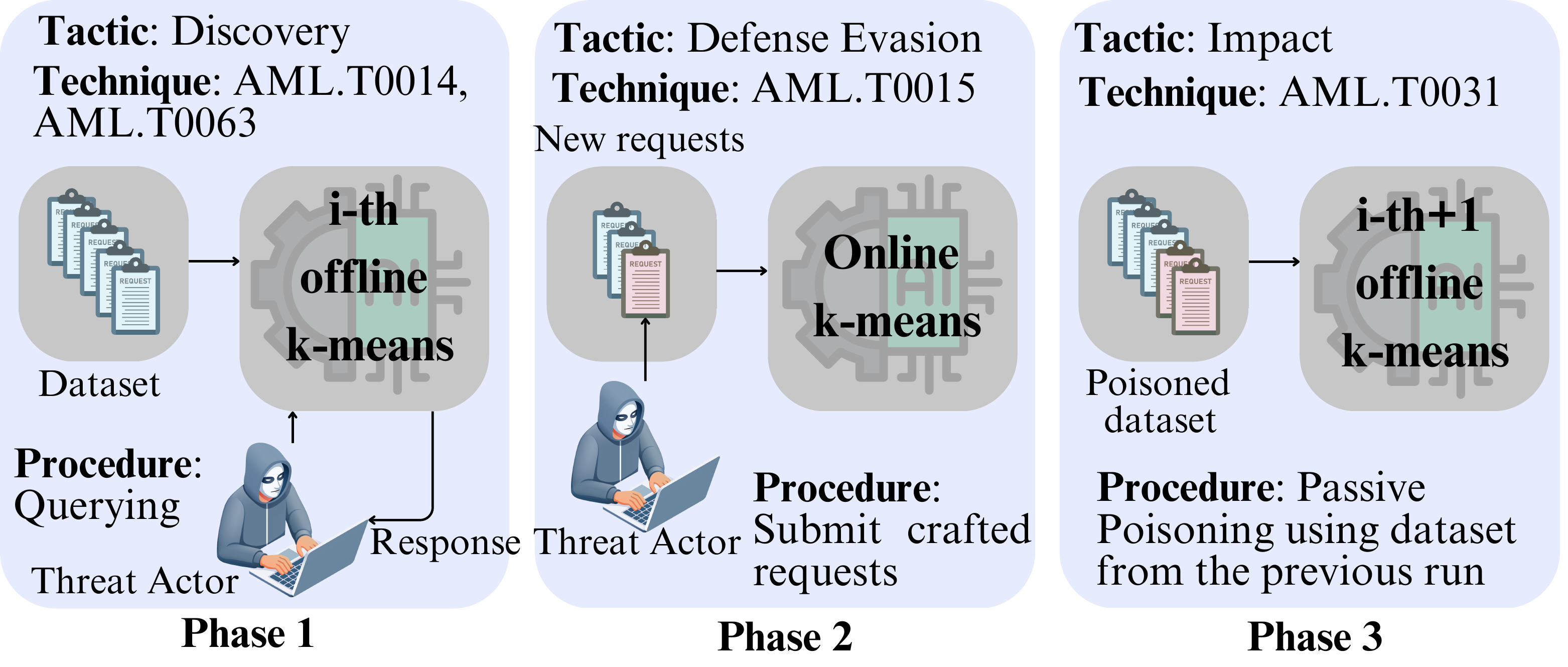}
    \caption{Mapping of the considered attacks to Attack-MITRE-ATLAS}
    \label{fig:attack-scenario}
\end{figure}

The threat actor begins with identifying the ML model family using AML.T0014 (discovery attack).
Once done, they access the model output through AML.T0063. 
The threat actor uses query-based probing to collect responses from the ML model and then refines their approach.
To initiate the causative attack, they bypass the offline k-means AD attacking the online phase (technique AML.T0015) using a defense evasion tactic. 
This allows passive poisoning of the $\mathbf{R} \cup \mathbf{R'}$ data, which affects the integrity of the model in the offline phase (technique AML.T0031).
Fig. \ref{fig:attack-scenario} illustrates the attack flow.
\subsection{Discovery (Exploratory Attack)}
After clustering all $\mathbf{r}_i \in \mathbf{R}$, the k-means algorithm produces the Voronoi diagram \cite{aurenhammer1991voronoi}.
For each centroid ${\bm \mu}_i \in \mathbf{R}$, where $i \in \{1,\ldots ,I\}$, its Voronoi cell (or cluster) $C_i$ is a polygon consisting of points that are closer to ${\bm \mu}_i$ than to any other centroid ${\bm \mu}_j$ for $j \neq i$.
The goal is to discover the clustering scheme inferring the decision boundaries represented by the set of points $\mathcal{X}_{ij}$ between adjacent $C_i$ and $C_j$ where $i \neq j$,
\begin{equation}
\label{equ:decision-boundary}
   \mathcal{X}_{ij} = \left\{\mathbf{x}_{ij} \in \mathbb{R}^5 \mid \bm{\omega}_{ij}^{\mathrm{T}} \cdot \mathbf{x}_{ij} + b_{ij} = 0 \right\}, 
\end{equation}
$\mathcal{X}_{ij} \neq \emptyset$, 
$\bm{\omega}_{ij} \in \mathbb{R}^{1\times5}$ is the normal vector to the decision boundary $\mathcal{X}_{ij}$, which determines the orientation of the separating hyperplane, and $b_{ij}$ is the scalar bias term that shifts the hyperplane. 
Let $\mathcal{Q}_{ij}=\{ \mathbf{q}_{ij_k} \} \subset \mathcal{X}_{ij}$, where $k\in \{ 1,\ldots , K\}$, be a subset of $K$ points on this hyperplane, which the threat actor aims to discover to approximate the decision boundary. 
Each such point $\mathbf{q}_{ij_k}$ should satisfy the hyperplane equation:
\begin{equation}
    \bm{\omega}_{ij}^{\mathrm{T}} \cdot \mathbf{q}_{ij_k} + b_{ij} = 0.
\end{equation}
To this end, the threat actor initiates an initial query set $\mathcal{Q}^0_{ij}=\{ \mathbf{q}^0_{ij_k} \} \subset \mathbb{R}^5$, consisting of $K$ queries (fake requests) sampled from the learned request domain, and being clustered in either $C_i$ or $C_j$.
Then, they identify clustering regions where the assigned cluster labels ($i$ and $j$) change (for $i$ to $j$ or vice versa) but their distance is small, i.e., they find $k, n \in \{1,\ldots , K\}$ such that:
\begin{equation}
\label{eq:changeoflabels}
    C_i=\mathcal{C}(\mathbf{q}^0_{ij_k}) \neq C_j=\mathcal{C}(\mathbf{q}^0_{ij_n}) \; \text{and} \; \|\mathbf{q}^0_{ij_k} - \mathbf{q}^0_{ij_n}\| \ll 1.
\end{equation}
Satisfying (\ref{eq:changeoflabels}) suggests the proximity of queries to a decision boundary $\mathcal{X}_{ij}$.
To approximate $\mathcal{X}_{ij}$, the threat actor refines the query set from $\mathcal{Q}^0_{ij}$ to $\mathcal{Q}_{ij}$ by minimizing the following objective function:
\begin{equation} 
    \label{equ:refinement-attack}
    \mathcal{L}_{\mathrm{ex}}\left(\mathcal{Q}_{ij}\right) = \sum_{k=1}^K  \left(\bm{\omega}_{ij}^{\mathrm{T}} \cdot \mathbf{q}_{ij_k} + b_{ij} \right)^2.
\end{equation}
To prevent excessive deviation of query points from their initial positions in $\mathcal{Q}^0_{ij}$, a regularization term is introduced to penalize large deviations:
\begin{equation}
    \label{equ:refinement-attack-regularized}
    \mathcal{L}_{\mathrm{ex}}\left(\mathcal{Q}_{ij} \right) = \sum_{k=1}^{K}  \left(\bm{\omega}_{ij}^{\mathrm{T}} \cdot  \mathbf{q}_{ij_k} + b_{ij} \right)^2 + \tau_{\mathrm{ex}} \sum_{k=1}^{K} \| \mathbf{q}_{ij_k}- \mathbf{q}_{ij_k}^0\|^2.
\end{equation}
Here, $\tau_{\mathrm{ex}}$ is the regularization parameter that balances decision boundary accuracy and closeness to the initial query set $\mathcal{Q}_{ij}^0$.
Using Stochastic Gradient Descent (SGD), the threat actor updates the query set $\mathcal{Q}_{ij}$ iteratively:
\begin{equation}
    \label{equ:sgd-attack-ex}
    \mathcal{Q}_{ij} \leftarrow \mathcal{Q}_{ij} - \eta_{\mathrm{ex}} \nabla \mathcal{L}_{\mathrm{ex}}\left(\mathcal{Q}_{ij} \right),
\end{equation}
where $\eta_{\mathrm{ex}}$ is the learning rate, and the gradient of the loss function with respect to the query set $\mathcal{Q}$ is given by:
\begin{multline}    
    \nabla \mathcal{L}_{\mathrm{ex}}\left(\mathcal{Q}_{ij} \right) = \\
    2 \sum_{k=1}^K \left( \left(\bm{\omega}_{ij}^{\mathrm{T}} \cdot \mathbf{q}_{ij_k} + b_{ij} \right) \bm{\omega}_{ij} + \tau_{\mathrm{ex}}\left(\mathbf{q}_{ij_k}-\mathbf{q}_{ij_k}^0 \right) \right).
\end{multline}
The threat actor inspects $\mathcal{X}_{ij}$ for the entire feature map, i.e., for all pairs of adjacent clusters $C_i$ and $C_j$, discovering all the boundary regions between them.

\subsection{Evasion Attack}
Once $\mathcal{Q}_{ij}$ are optimized for all $\mathcal{X}_{ij}$, an attacker aims to force clustering $\mathcal{C}$ to wrongly cluster benign workloads $\mathbf{R}_{\mathrm{s}}'=\{\mathbf{r}_{\mathrm{s}_1},\ldots \mathbf{r}_{\mathrm{s}_{M_{\mathrm{s}}'}}\}\subset \mathbf{R'}$ ($M_{\mathrm{s}}'$ being the cardinality of $\mathbf{R}_{\mathrm{s}}'$) by generating perturbations $\bm{\Gamma} = \{\bm{\gamma}_1, \ldots \bm{\gamma}_{M_{\mathrm{s}}'}\} \in \mathbb{R}^5$:
\begin{equation}
\label{equ:perturbed}
    \forall m \in \{1,\ldots,M_{\mathrm{s}}'\}: \mathbf{r}_{\mathrm{p}_m} = \mathbf{r}_{\mathrm{s}_m} + \bm{\gamma}_m.
\end{equation}
For a workload $\mathbf{r}_{\mathrm{s}_m}$ where $ m\in\{1,\ldots,M_{\mathrm{s}}'\}$, the threat actor first identifies a cluster $C_i$ (based on clustering $\mathcal{C}$) with centroid $\bm{\mu}_i$, as well as its neighboring target cluster $C_j$ with centroid $\bm{\mu_j}=\left[ \mu_j^{(\text{CPU})},\mu_j^{(\text{IO})}, \mu_j^{(\text{E2E})}, \mu_j^{(\text{SIZE})}, \mu_j^{(\text{PER})}\right] $ such that:
\begin{equation}
    \exists \phi \in \{\text{CPU}, \text{IO}, \text{SIZE}\}\mid\mu_j^{(\phi)} < \mu_i^{(\phi)},
\end{equation}
To overload the target cluster, they optimize $\bm{\gamma}_m$ to minimize the distance between $\mathbf{r}_{\mathrm{p}_m}$ and $\hat{\bm{\mu}}_i$ while maximizing the distance from $\bm{\mu}_i$:
\begin{equation}
    \label{equ:refinement-attack-evasion}
    \mathcal{L}_{\mathrm{ev}}(\gamma_i) = \|\mathbf{r}_{\mathrm{s}_m}+\bm{\gamma}_m-\hat{\bm{\mu}}_i\|^2 - \tau_{\mathrm{ev}}\|\mathbf{r}_{\mathrm{s}_m}+\bm{\gamma}_m-\bm{\mu}_i\|^2,
\end{equation}
where $\tau_{\mathrm{ev}}$ is a weighting parameter.
 
For optimization, the threat actor applies the Projected Gradient Descent (PGD) to update $\gamma_i$. 
The update rule for $\bm{\gamma}_m$ is:
\begin{equation}
    \label{equ:pgd-evasion}
    \bm{\gamma}_m \leftarrow \bm{\gamma}_m - \eta_{\mathrm{ev}} \nabla \mathcal{L}_{\mathrm{ev}}(\bm{\gamma}_m),
\end{equation}
where $\eta_{\mathrm{ev}}$ is the learning rate and $\nabla \mathcal{L}_{\mathrm{ev}}(\bm{\gamma}_m)$ is the gradient of the loss function with respect to $\bm{\gamma}_m$:
\begin{equation}
    \nabla \mathcal{L}_{\mathrm{ev}}(\bm{\gamma}_m) = 2(\mathbf{r}_{\mathrm{s}_m}+\bm{\gamma}_m-\hat{\bm{\mu}}_i) - 2\tau_{\mathrm{ev}}(\mathbf{r}_{\mathrm{s}_m}+\bm{\gamma}_m-\bm{\mu}_i).
\end{equation}
To ensure that the perturbation remains within a specified distance from the benign workload, $\bm{\gamma}_m$ is projected back into the feasible region using the Euclidean projector operator, i.e., $\bm{\gamma}_m \leftarrow \Pi_{\Gamma}(\bm{\gamma}_m)$.  
The projection operator $\Pi_{\Gamma}$ for an $\ell_2$-ball constraint with radius that is the maximum allowable cluster overloading $\epsilon_{\mathrm{ev}}$ is defined as:
\begin{equation}
\Pi_{\Gamma}(\bm{\gamma}_m) = 
\begin{cases}
\bm{\gamma}_m, & \text{if } \|\bm{\gamma}_m\|_2 \leq \epsilon_{\mathrm{ev}}, \\
\epsilon_{\mathrm{ev}} \frac{\bm{\gamma}_m}{\|\bm{\gamma}_m\|_2} & \text{if } \|\bm{\gamma}_m\|_2 > \epsilon_{\mathrm{ev}}.
\end{cases}
\end{equation}
\section{The Proactive Hardening}
\label{sec:defense}
The goal of adversarial training is to make clustering $\mathcal{C}$ robust to perturbations, such that the predicted labels of clustered requests do not change despite 
the deliberately introduced perturbations $\tilde{\bm{\Gamma}}=\{\tilde{\bm{\gamma}}_1, \ldots \tilde{\bm{\gamma}}_{\tilde{M}}\} \in \mathbb{R}^5$: 
\begin{equation}
     \forall m\in\{1,\ldots,\tilde{M}\}:\; \arg\mathcal{C}(\mathbf{r}_m) = \arg \mathcal{C}\left(\mathbf{r}_m+\tilde{\bm{\gamma}}_m\right).
\end{equation}
The adversarial training formulates an optimization problem to minimize the likelihood of misclassification under $|\tilde{\bm{\gamma}}_m| \leq \iota$, with $\iota$ being the maximum perturbation norm allowed. 
The (deliberate) perturbation generation method proposed for the adversarial training is PGD.
The objective for generating the perturbation is to find:
\begin{align} 
\label{equ} 
    \tilde{\bm{\gamma}}_m \leftarrow \arg \max_{\tilde{\bm{\gamma}}_m} \mathcal{L}_{\mathrm{at}}(\mathcal{C}\left(\mathbf{r}_m+\tilde{\bm{\gamma}}_m\right), i)\;\mathrm{s.t.}\; |\tilde{\bm{\gamma}}_m| \leq \iota \mathrm{,}\\
    \mathcal{L}_{\mathrm{at}} = -\sum_{i=1}^I i\log(\mathcal{C}(\mathbf{r}_m+\tilde{\bm{\gamma}}_m)),
\end{align}
where the cross-entropy loss $\mathcal{L}_{\mathrm{at}}$ measures the discrepancy between the predicted output of the model for the perturbed workload $\mathbf{r}_m + \tilde{\bm{\gamma}}_m$ and the true label $i$.
The PGD update rule at each iteration for a perturbation $\tilde{\bm{\gamma}}_m$ is:
\begin{equation}
\tilde{\bm{\gamma}}_m \leftarrow \Pi_{\tilde{\bm{\Gamma}}} \left( \tilde{\bm{\gamma}}_m + \eta_{\mathrm{at}} \frac{\partial \mathcal{L}_{\mathrm{at}}}{\partial \tilde{\bm{\gamma}}_m} \right),
\end{equation}
or equivalently:
\begin{equation}
\tilde{\bm{\gamma}}_m \leftarrow \Pi_{\tilde{\bm{\Gamma}}} \left( \tilde{\bm{\gamma}}_m + \eta_{\mathrm{at}} \nabla_{\tilde{\gamma}_i} \mathcal{L}_{\mathrm{at}}(\mathcal{C}(\mathbf{r}_m + \tilde{\bm{\gamma}}_m), i) \right),
\end{equation}
where $\eta_{\mathrm{at}}$ is the learning rate, and $\Pi_{\tilde{\bm{\Gamma}}}$ denotes the projection operator that ensures $\tilde{\bm{\gamma}}_m$ remains within the allowed perturbation norm $\epsilon_{\mathrm{at}}$.
Specifically, this projection can be defined as:
\begin{equation}
    \Pi_{\tilde{\Gamma}}(\tilde{\bm{\gamma}}_m) = \begin{cases}
\tilde{\bm{\gamma}}_m & \text{if } \|\tilde{\bm{\gamma}}_m\| \leq \epsilon_{\mathrm{at}}, \\
\epsilon_{\mathrm{at}}\frac{\tilde{\bm{\gamma}}_m}{\|\tilde{\bm{\gamma}}_m\|} & \text{if } \|\tilde{\bm{\gamma}}_m\| > \epsilon_{\mathrm{at}}.
\end{cases}
\end{equation}
\section{Simulation Results}
\label{sec:sim}
We have examined the considered fog network scenario with RPS based on request clustering and network resource matching by the dispatcher by computer simulations using Python. We have examined attack consequences and proactive hardening. The VM-fixed and dispatcher-based configurations have been assumed for attack-free and attacked network scenarios.
Workload requests $\mathbf{R}$ and VM offers were created with a range of feature values as detailed in \cite{kopras2023communication}.
Since the features do not share the same range, to avoid bias, the features of $\mathbf{R}$ are normalized before processing.
Two metrics are considered for evaluation, namely resource utilization (RU) and task drop ratio (TD).
RU represents the ratio of the utilized VM resource to the allocated VM resources among all VMs:

\begin{equation}
\label{equ:RU}
    \text{RU}_\% = \frac{100}{4}\sum_{m=1}^{M}\sum_{i=1}^I\left(\frac{r_m^{(\text{CPU})}}{o_i^{(\text{CPU})}}+ \frac{r_m^{(\text{IO})}}{o_i^{(\text{IO})}} + \frac{r_m^{(\text{SIZE})}}{o_i^{(\text{CAP})}} +  \frac{r_m^{({\frac{E_b}{N_0}})}}{o_j^{({\frac{E_b}{N_0}})}}\right),
\end{equation}
where $r_m^{({\frac{E_b}{N_0}})}$ and $ o_i^{({\frac{E_b}{N_0}})}$ are values calculated for requested and offered PER respectively, and they are obtained from the low-density parity check LDPC coded channel mapping of PER to the average energy per bit  $\text{PER} \leftrightarrow \frac{E_b}{N_0}$ for Nakagami channel.

TD quantifies the number of workloads served 
after their deadlines or suffering from 
higher 
PER than their tolerated one.
\begin{equation}
\label{equ:GR}
    \text{TD}_\% = \frac{|\mathbf{r}_m \rightarrow \bm{\mu}_i: t_i > r^{(\text{E2E})}_m \vee  e_i> r^{(\text{PER})}_m|}{|\mathbf{R}|} \cdot 100,
\end{equation}
where the completion time $t_i$ is the summation of the serving time and the transmission time. 
In the notation above, for simplification, we omitted the index indicating the number of submitted request butches.
The first-come-first-serve CPU/IO burst cycle scheduling is performed after provisioning.
Thus, Kendall's notation of the given system is $M/G/I/\infty/\infty/G$.
The transmission time refers to the uplink propagation delay.

The parameters used in our simulations are listed in Table \ref{tab:simulation_parameters}.
\begin{table}[htbp]
    \caption{Simulation Parameters}
    \centering
    \begin{tabular}{|c|c||c|c|}
    \hline
    \textbf{Parameter}                       & \textbf{Value} & \textbf{Parameter}                                           & \textbf{Value}   \\ \hline
    $l$                                      & 3              & $|\mathcal{Q}_{ij}|$                                         & 10               \\ \hline
    $M$                                      & [100, 1000]    & $\zeta_1$                                                    & 9/10             \\ \hline
    $M'$                                     & [100, 1000]    & $\zeta_2$                                                    & 1/10             \\ \hline
    ${\alpha_i}/{\tilde{\alpha_i}}$          & [1, 3]         & $\eta_{\mathrm{ex}}, \eta_{\mathrm{ev}}, \eta_{\mathrm{at}}$ & 1/1000           \\ \hline
    $\tau_{\mathrm{ex}}, \tau_{\mathrm{ev}}$ & 1/2            & $\epsilon_{\mathrm{ev}}, \epsilon_{\mathrm{at}}$             & 4/100            \\ \hline
    $\iota$                                  & 1/1000         & PER range                                                    & \cite{5QI}       \\ \hline
    \end{tabular}
    \label{tab:simulation_parameters}
\end{table}
The real-time k-means initialized model converges after only 4 iterations (even for a  high number of clusters) compared to 11 iterations for the non-initialized model.
In addition, the average deviation of the final centroids from the initial ones is found to be 8\%.
These results indicate the fast and correct convergence of the adapted real-time clustering model with the pre-defined initial centroids.

\begin{figure}[thb]
    \centering
    \includegraphics[width=0.9\columnwidth]{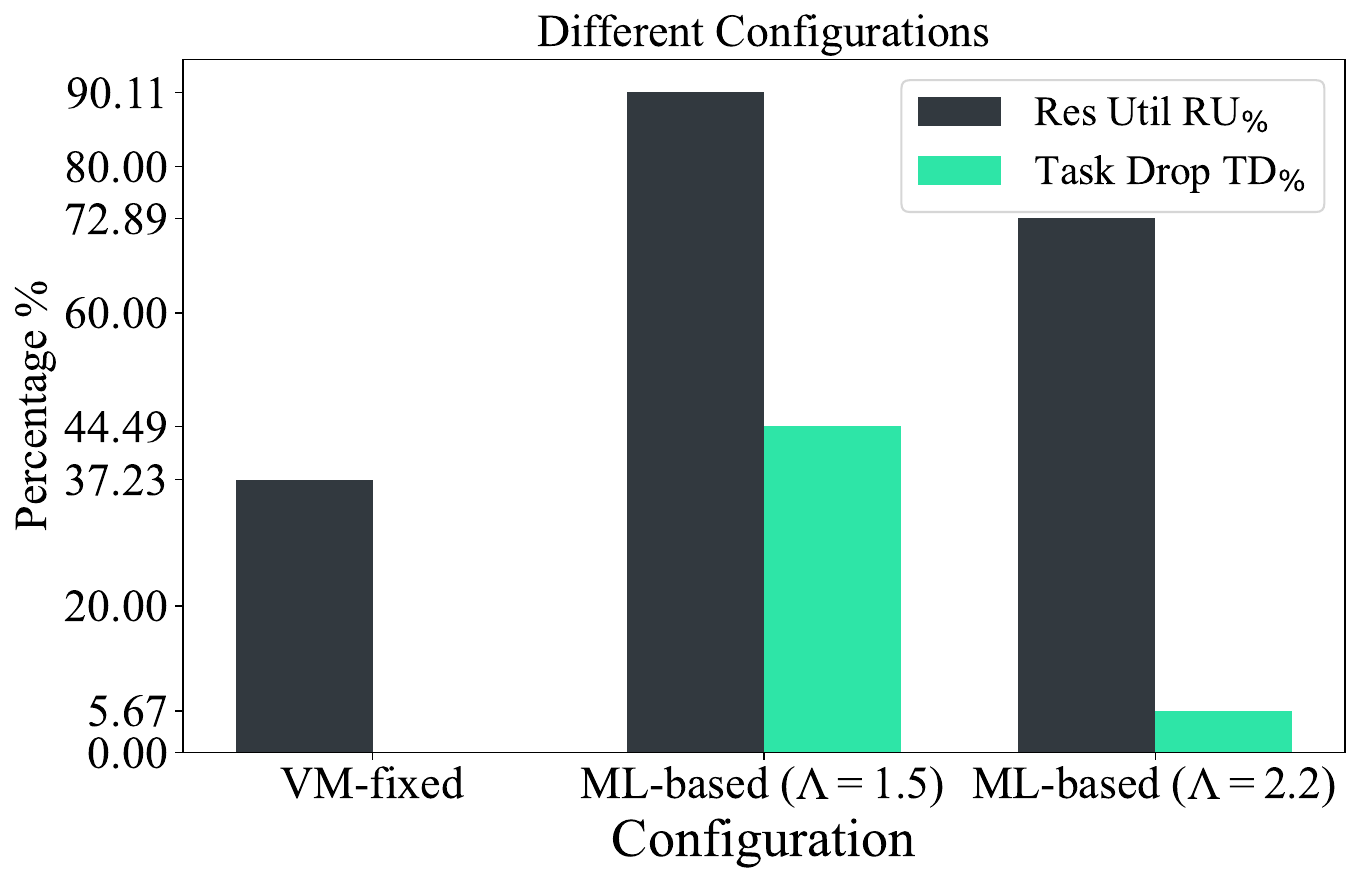}
    \caption{Simulation results for correct resource provisioning}
    \label{fig:Legitimate-Results}
\end{figure}

Fig. \ref{fig:Legitimate-Results} depicts RU for different configurations in the legitimate scenario.
One can observe that the clustering-based RPS improves RU compared to the 
optimal VM-fixed approach (described in \cite{alnoman2020machine}). 
Our dispatcher-based approach reaches 90\% RU which outperforms VM-fixed one by approximately 240\%.
This result is obtained for $\Lambda=2.2$, which makes the RPS suffer from high TD (45\%).
To ensure comparable TD to the optimal VM-based with a high RU, $\Lambda$ was decreased to 1.5, which provides a 71.35\% RU and 6\% TD.
This shows the trade-off between optimal RU and low TD ratio and gives an insight into the $\Lambda$ value that can be optimized.

\begin{figure}[bht]
    \centering
    \includegraphics[width=0.9\columnwidth]{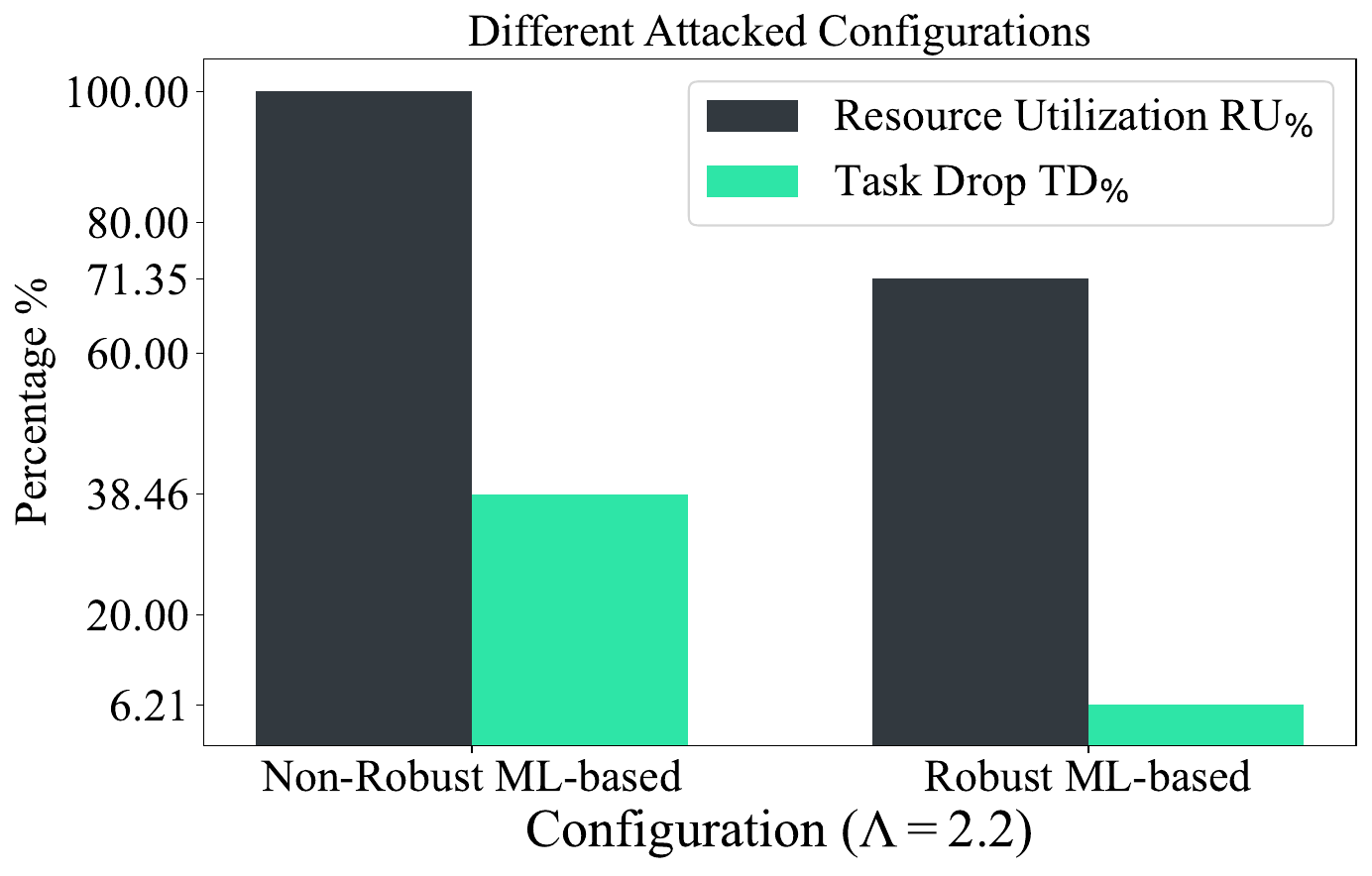}
    \caption{Attacked vs. protected ML-based provisioning simulation results} 
    \label{fig:Attacked-Results}
\end{figure}

In the attack scenario (Fig. \ref{fig:Attacked-Results}), non-robust RPS experiences a surge RU reaching 100\%. 
This increase is attributed to the excessive load on the VMs, as the attack redirects high-demand workloads to clusters with low resources.
The resulting overload significantly increases the completion time of the altered workloads, leading to TD rise to 38\%.
These results show that the threat actor succeeded in manipulating $\Lambda$.

However, introducing robustness to the ML-based provisioning successfully recovers its performance.
The RU achieves approximately 98\% of its original value of the dispatcher configuration.
Furthermore, the TD value drops to 6\%, which is fairly comparable to the value of the no-attack scenario.
It should be noted that data distribution helped the threat actor achieve their goal. 
The data is uniformly distributed and the clusters are very close to each other; thus, it is more probable that the data manipulation has more impact on the system than the distinct clusters. 
Moreover, the high diversity of the clustering features increases their number, which gives the threat actor more boundaries to attack.

\subsection*{Proactive hardening complexity}
The total complexity of adversarial training is obtained as:
\begin{equation}
    \mathcal{O}(|\tilde{\Gamma}| \cdot \dim(\tilde{\Gamma}) \cdot I \cdot I_{\text{PGD}}),
\end{equation}
where $|\tilde{\Gamma}|$, $\dim|\tilde{\Gamma}|$, and $I_{\text{PGD}}$ are the number of perturbations used in the adversarial training, the dimensionality of the perturbations (here equal 5, as the dimension of requests), and the number of PGD iterations, respectively.
As the number of clusters $I$ and the dimensionality  $\dim(\Bar{\Gamma})$ increase, the computational overhead scales linearly, emphasizing the importance of optimizing these parameters for scalability. 
In real-time applications, the delay added by adversarial training is of  critical consideration, quantified by 
\begin{equation}t_{\text{total}} = |\tilde{\bm{\Gamma}}| \cdot t_{\mathrm{c}} + |\tilde{\bm{\Gamma}}| \cdot I_{\text{PGD}} \cdot t_{\text{grad}},
\end{equation} 
where $t_{\mathrm{c}}$ is the classification time and $t_{\text{grad}}$ is the gradient computation time. 
A distributed computing mechanism is developed by partitioning the workloads across the $I$ parallel nodes, reducing the effective batch size per node to $|\tilde{\bm{\Gamma}}|/I$.

\section{Conclusion}
\label{sec:conc}
Above, an ML-based algorithm was proposed for allocating 2C workload to VMs in a low-complexity real-time application scenario.
The results show a faster and correct convergence for the initialized k-means model.
Next, a discovery and an evasion attack were considered that were launched against the k-means classifier to affect its integrity.
Simulation results prove that the threat actor successfully degraded resource provisioning performance by overloading the VMs.
To mitigate this, a new adversarial training method was proposed to introduce robustness to the ML classifier.
Computer simulation results of the robust classifier applying our method show a near-optimal defending capability counteracting attacks launched, recovering the performance of the resource provisioning system.


\bibliographystyle{IEEEtran}
\bibliography{bibliography.bib}
\end{document}